\documentclass[a4paper,fleqn,usenatbib]{mnras}

\usepackage{newtxtext,newtxmath}
\usepackage[T1]{fontenc}
\usepackage{ae,aecompl}
\usepackage{booktabs}

\usepackage{graphicx}	
\usepackage{amsmath}	
\usepackage{amssymb}	

\title[Constraints on proto-planetary disc evolution]{
Constraining proto-planetary disc evolution using accretion rate and disc mass measurements: the
usefulness of the dimensionless accretion parameter
}

\author[Rosotti et al.]{
Giovanni P. Rosotti,$^{1}$\thanks{E-mail: rosotti@ast.cam.ac.uk}
Cathie J. Clarke,$^{1}$
Carlo F. Manara,$^{2}$
Stefano Facchini$^{3}$
\\
$^{1}$Institute of Astronomy, University of Cambridge, Madingley Road, Cambridge, CB3 0HA, UK\\
$^{2}$Scientific Support Office, Directorate of Science, European Space Research and Technology Centre (ESA/ESTEC),\\ Keplerlaan 1, 2201 AZ, Noordwijk, The Netherlands\\
$^{3}$Max-Planck-Institut f\"{u}r Extraterrestrische Physik, Giessenbachstrasse 1, 85748 Garching, Germany
}

\date{Accepted 2017 March 08. Received 2017 March 04; in original form 2016 July 22}

\pubyear{2017}

\begin{document}
\label{firstpage}
\pagerange{\pageref{firstpage}--\pageref{lastpage}}
\maketitle

\begin{abstract}

We explore how measurements of protoplanetary disc masses and accretion rates provided by surveys of star forming regions can be analysed via the {\it dimensionless accretion parameter}, which we define as the product of the accretion rate and stellar age divided by the disc mass. By extending and generalising the study of Jones et al (2012), we demonstrate that this parameter should be less than or of order unity for a wide range of evolutionary scenarios, rising above unity only during the final stages of outside in clearing by external photoevaporation. We use this result to assess the reliability of disc mass estimates derived from CO isotopologues and submm continuum emission by examining the distribution of accretion efficiencies in regions which are not subject to external photoevaporation. We find that while dust based mass estimates produce results compatible with theoretical expectations assuming canonical dust/gas ratio, the systematically lower CO based estimates yield accretion efficiencies significantly above unity in contrast with the theory. This finding provides additional evidence that CO based disc masses are an under-estimate, in line with arguments that have been made on the basis of chemical modelling of relatively small samples. On the other hand, we demonstrate that dust based mass estimates are sufficiently accurate to reveal distinctly higher accretion efficiencies in the Trapezium cluster, where this result is expected given the evident importance of external photoevaporation. We therefore propose the dimensionless accretion parameter as a new diagnostic of external photoevaporation in other star forming regions.

\end{abstract}




\section{Introduction}

\label{sec:intro}


One of the main scientific motivations for the construction of facilities that can conduct surveys of large samples of protoplanetary discs is that the statistical distributions of disc properties
can be used to place constraints on evolutionary models. In turn, these are needed to characterise the formation environments of planets and build a successful planet formation theory. This applies {\it par excellence} to
observatories such as ALMA where disc surveys in submm continuum and emission lines from the isotopologues of
CO \citep{Ansdell2016,Barenfeld2016,Pascucci2016} yield estimates of disc masses \citep{WilliamsBest2014,Miotello2016,Boneberg2016}; in parallel, the possibility of simultaneous
spectroscopic characterisation across a wide range of near ultraviolet to near infrared
wavelengths with instruments such as X-shooter provides an opportunity to
assemble data on disc accretion rates of unprecedented accuracy (\citealt{Manara2016Cha,Alcala2014,Alcala2016}; see also \citealt{HerczegHillenbrand2008,Ingleby2013}).

There are however several inter-woven difficulties in straightforwardly using such
data for constraining disc evolution models. There is no single family of disc evolution models that can be `calibrated' against observational data: the models contain a number of uncertainties concerning the mechanism for angular momentum
removal/redistribution and its associated timescales \citep{TurnerPPVI}, combined with further
uncertainties associated with the roles of internal/external photoevaporation \citep{Alexander2014}.
 At the same time, also, observational data
is in need of calibration because of uncertainties in relating observed quantities to disc properties. This problem is particularly acute for dust and CO isotopologues because of their varying abundance with respects to the dominant mass component \citep{Favre2013,McClure2016}.

At first sight then, the problem of ill constrained theoretical models combined
with uncertain observational calibration limits the utility of
observational survey data. In this paper we present an analysis path which
assesses the available data in relation {\it both} to its intrinsic reliability
{\it and} to its ability to discern environmental differences between different star forming regions. Following \citet{Jones2012}, the crux of the argument
is based on comparing disc masses with the product of accretion rate
and age. We
assess the much larger observational dataset currently available and also
generalise the theoretical models considered. For ease of discussion we introduce some new terminology, defining
the {\it dimensionless accretion parameter} as
\begin{equation}
\eta = \tau \dot{M} /M_\mathrm{disc}
\label{eq:eta}
\end{equation} 
and comparing observational estimates
of this quantity with model predictions. In this definition a system with high dimensionless accretion parameter is one for which the accretion rate is high
given its age and disc mass. We will show that self-regulation
of disc evolutionary models implies that a high dimensionless accretion parameter
{\it cannot} be achieved by simply invoking more rapid angular momentum
transport within discs and show that it instead points to the role of external photoevaporation.

In outline
we consider the properties of the whole family of possible theoretical models
and use this to argue that the data in the literature generically favours the use of
sub-mm dust continuum data (rather than CO isotopologue data) as a measure of disc mass.
With this assumption we then further examine whether the existing data is \textit{accurate} enough to find possible differences between star forming regions. In particular we consider the central region of the Orion Nebula Cluster (ONC) where external photo-evaporation is expected to be important and demonstrate that indeed, by comparison with low-mass star forming regions, the available
data bears an imprint of the role of external photoevaporation in setting
disc properties.

This paper is structured as follows. In section \ref{sec:expectations} we present theoretical arguments to illustrate the utility of the dimensionless accretion parameter previously mentioned. In Section \ref{sec:observations} we analyse the existing data in light of the dimensionless accretion parameter  we have introduced. Finally in Section \ref{sec:conclusions} we draw our conclusions.

\section{Theoretical expectations}
\label{sec:expectations}
\subsection{Viscous evolution}
\label{sec:viscous}

We here consider discs that evolve as a result of what we loosely term viscous processes, i.e. 
processes that \textit{redistribute} angular momentum
within the disc, leading to radial (accretion) flows 
of velocity $v_r(r)$.  Such processes are commonly invoked to explain the empirical evidence that young stellar objects accrete. Although the exact mechanism responsible for their existence is not currently known, the best candidate is the Magneto Rotational Instability (MRI); see \citet{TurnerPPVI} for a review. 

In the absence
of other processes, viscous processes attempt to set up
a quasi-steady state in which the local accretion rate $\dot M(r)$ is
independent of radius over much of the disc. Defining the local viscous
timescale as $t_\nu (r) = r/v_r$, and writing $\dot M(r) = 2 \pi \Sigma r v_r$
(where $\Sigma$ is the local surface density in the disc) we can then
write $\dot M(r) = 2 \pi \Sigma r^2/t_\nu (r)$. Parametrising $\Sigma$ as a radial power law 
$\Sigma \propto r^{-p}$, where observationally
$p$ is measured in the range $0.5$ to $1.5$ \citep{WilliamsCieza}, we conclude that $\Sigma r^2$ is
an increasing function of $r$. Thus, if observed discs have been
significantly shaped by viscous evolution,  $t_\nu$ is an increasing function\footnote{Note that this is equivalent to requiring that most of the disc mass is at large radii.}
of $r$. In this case, the quasi-steady state has two additional properties:
\begin{enumerate}
\item the outer disc radius is set by the requirement that $t_\nu$ in the outer
disc is of order  the system age $\tau$ (i.e. the disc has just had time to grow
by viscous evolution to its present size);
\item since most of the disc
mass is at large radii, $M_\mathrm{disc} (r) \sim \pi \Sigma(r) r^2$ and the quasi-steady
state condition can be written in the form $ \dot M = \sigma M_\mathrm{disc}(r)/t_\nu (r)$, where $\sigma$ is a constant of order unity which depends on the
surface density profile in the quasi-steady state.
\end{enumerate}
Taken together, the two properties imply that the viscous quasi-steady state is characterised by the
condition $\dot M \sim M_\mathrm{disc}/\tau,$ i.e. $\eta = \tau \dot M / M_\mathrm{disc} \sim 1$. We thus assign a dimensionless accretion parameter of unity to a disc that is accreting according to this expectation. Note how this establishes a connection between the mass accretion rate \textit{onto the star} and the \textit{outer} parts of the disc (typically at hundreds of au) where most of the mass resides.

 The above qualitative description is borne out by many
numerical calculations including those of \citet{Jones2012} and those presented
below; it can also be demonstrated analytically in the case of discs
with a simple power law parametrisation of viscosity on radius, which
evolve according to viscous similarity solutions \citep{LyndenBellPringle74,Hartmann98}. It is however worth emphasising that
the features noted above (in particular the prediction $\eta \sim 1$) 
are {\it not}  dependent on this restrictive assumption
and are a generic feature of discs that have attained  a viscous
quasi-steady state. We perform a series of numerical experiments to demonstrate this in Section \ref{sec:visc_exp} through using some
extreme parameter choices for the viscous properties of the disc. In Section \ref{sec:other} we investigate instead what happens when a viscous quasi steady-state \textit{cannot} be attained.

\subsection{Method}
\label{sec:method}

We solve the evolution equation of the disc surface density $\Sigma (R,t)$ (e.g. \citealt{LyndenBellPringle74,Pringle81}) due to the kinematical viscosity $\nu(\Sigma, R)$:
\begin{equation} \label{eq:sigma_evol}
\frac{\partial \Sigma}{\partial t} = \frac{1}{R}\frac{\partial}{\partial R}\left[ 3R^{1/2} \frac{\partial}{\partial R}\left(\nu \Sigma R^{1/2}\right) \right] - \dot{\Sigma} (R),
\end{equation}
where $\dot{\Sigma} (R)$ represents a potential loss-term to model the effects of external photo-evaporation.

The equation is discretised on a grid of points uniformly spaced in $R^{1/2}$ and then integrated in time using a standard explicit finite-difference method second order accurate in space and first order accurate in time \citep{Pringle86}. When taking into account external photo-evaporation, we follow \citet{Clarke2007} by removing mass from the outer edge of the disc. Namely, we locate the outer edge of the disc by finding the grid cell where the column density is closest to some low constant value (provided that this value is sufficiently low, the results are insensitive to the exact value due to the steepness of the surface density profile close to the outer edge; in our calculations we employ a value of $10^{-8}$ g cm$^{-2}$). We then progress inwards removing a total mass of $\dot{M}_{\rm ph} \Delta t$, where $\Delta t$ is the length of the timestep and $\dot{M}_{\rm ph}$ the total mass-loss due to external photo-evaporation, which is assumed \citep[cf.][]{Adams2004,Facchini2016} to be concentrated at the disc outer edge. Other works \citep{MitchellStewart2010,Anderson2013,Kalyaan2015} have taken a similar approach. See \citet{Clarke2007} for the photo-evaporation rates that we assume, which are parametrised from \citet{Adams2004} and \citet{Johnstone1998} in the super-critical regime. They correspond to a value of the interstellar FUV field $G_0=3000$ \citep{Habing68} for a $1 \ M_\odot$ star.

\subsection{Sensitivity of results to radial dependence of the viscosity}
\label{sec:visc_exp}

\begin{figure*}
\includegraphics[width=\columnwidth]{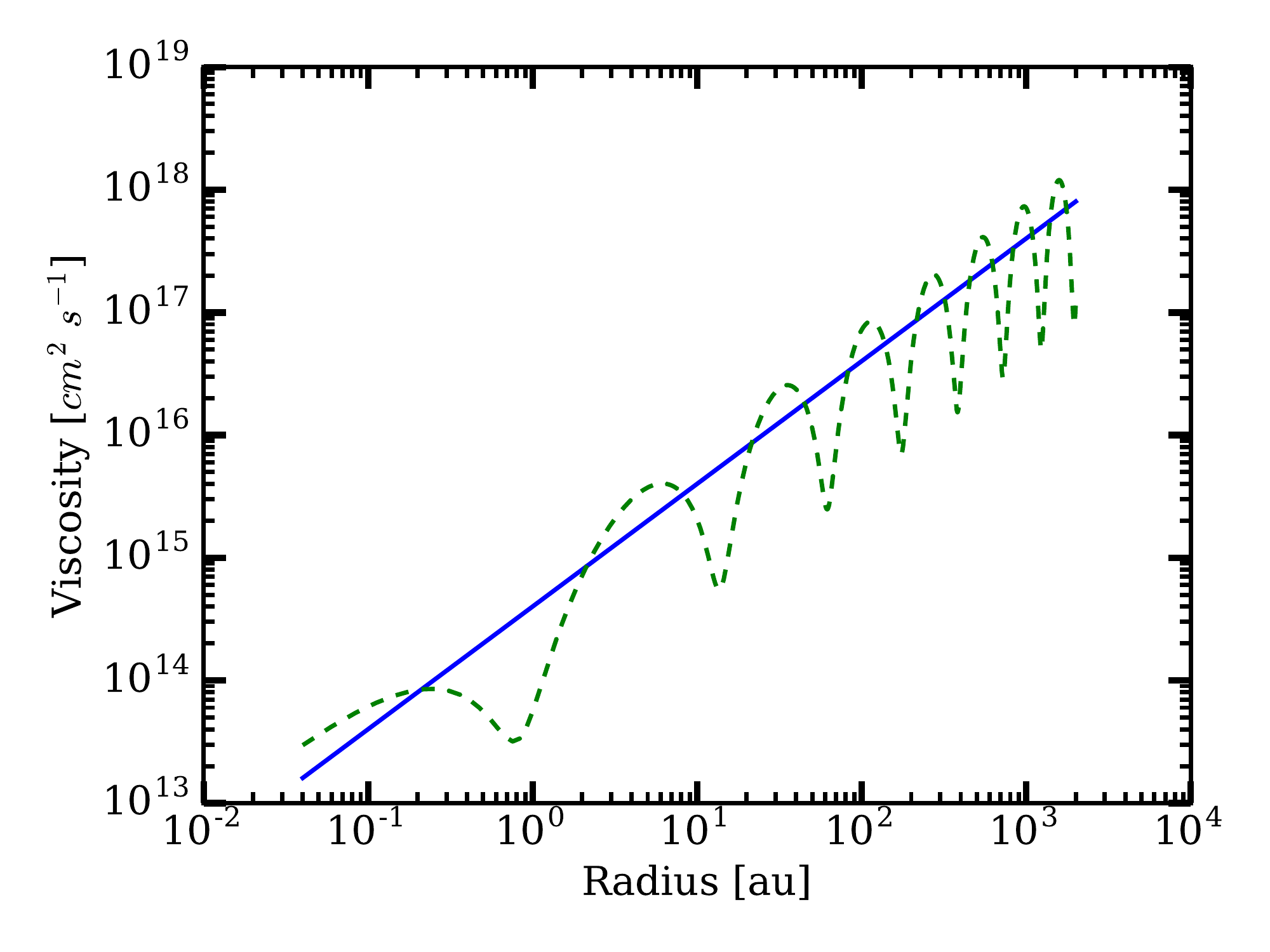}
\includegraphics[width=\columnwidth]{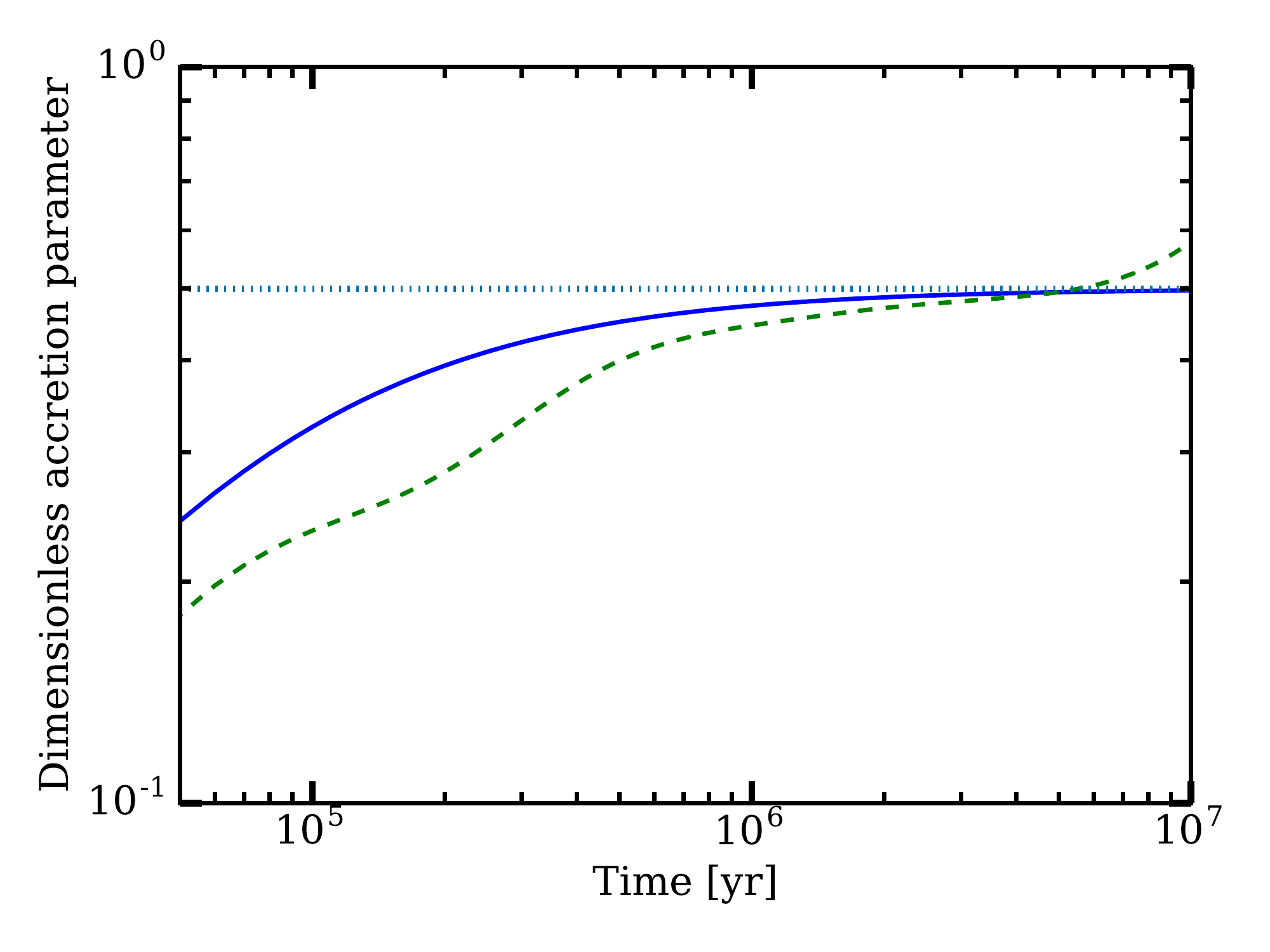}

\caption{Left panel: the green dashed line shows the viscosity profile applied; for reference we plot also the background law  $\nu \propto r$ (blue solid line). Right panel: evolution in time of the dimensionless accretion parameter for the two cases (the colors have the same meaning of the previous plot). For reference the horizontal blue dotted line shows a dimensionless accretion parameter of 0.5. Despite the extreme substructure we artificially imposed on the viscosity, there is very little difference in the evolution of the dimensionless accretion parameter.}
\label{fig:varying}
\end{figure*}

In what follows we use as initial conditions a self-similar solution \citep{LyndenBellPringle74} with $\gamma=1$, an initial radius of 10 au and an initial disc mass of $0.05 \ M_\odot$. The value of $\alpha$ \citep{ShakuraSunyaev73} is 0.01 (if not specified otherwise) and we assume a flaring disc with $H/R=0.03$ at 1 au.


In this section we present some numerical experiments which have the goal of illustrating the claims made previously. In particular, we show that $\eta \sim 1$ is reached even if there are radial variations in the efficiency of
viscous transport. In Figure \ref{fig:varying} we illustrate the highly artificial case
that the viscosity is subject to large amplitude radial fluctuations
superposed on a background law $\nu \propto r$ (left panel; the blue solid line is the background and the green dashed the oscillatory profile). The right
hand panel illustrates that there is very little difference in the
evolution of the dimensionless accretion parameter as a function of time: both models settle
into the state where $\eta \sim 1$ once the system age exceeds
the maximum viscous timescale in the disc (which is at large radii in 
both cases). Note that for $\gamma=1$, $\sigma$ equals 0.5 (see Section \ref{sec:viscous}), which explains why the accretion parameter remains slightly smaller than unity.

\begin{figure}
\includegraphics[width=\columnwidth]{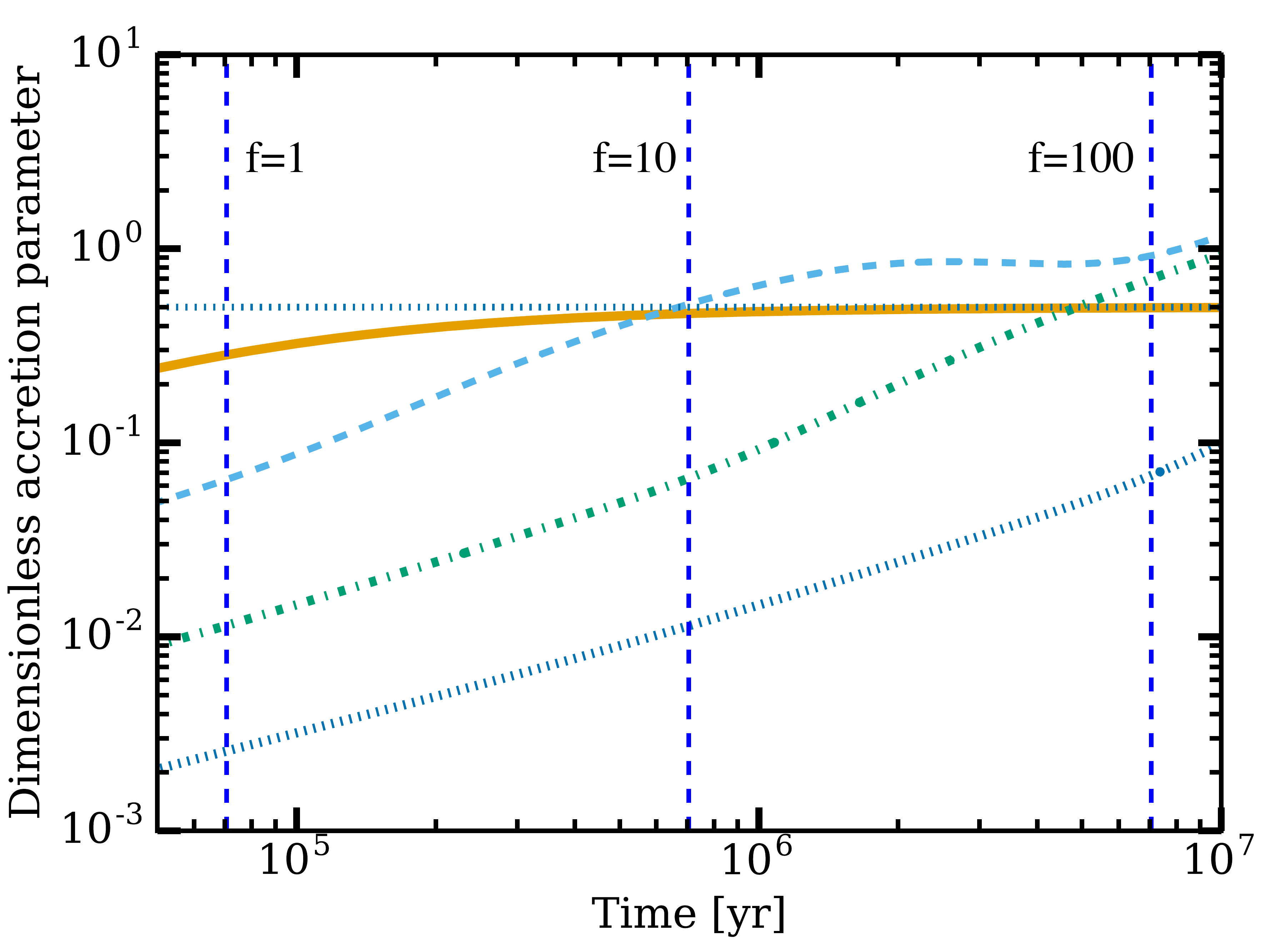}
\caption{Evolution in time of the dimensionless accretion parameter for the case of a dead zone between 1 and 10 au. The different lines correspond to different values of the parameter $f$ (see text): 1 (orange solid), 10 (light blue dashed), 100 (green dotted-dashed) and 1000 (blue dotted). The dashed blue vertical lines represent the viscous timescale at the outer edge of the dead zone (i.e. 10 au) for the different values of f (note that the case of f=1000 is outside the scale of the plot). The blue dotted horizontal line shows a reference dimensionless accretion parameter of 0.5. For f=1 and 10 the dimensionless accretion parameter is not significantly altered. For f=100 the dimensionless accretion parameter reaches unity only after several Myr; finally for f=1000 the dimensionless accretion parameter is always lower than unity. }
\label{fig:dead}
\end{figure}

In Figure \ref{fig:dead} we show a more realistic case where the viscosity
is dropped by a factor $f$ at radii between 1 and 10 au, possibly corresponding to
the case of a region of weak magneto-turbulence, a so-called dead zone (see \citealt{Armitage2011, TurnerPPVI} for reviews of the topic). We test values of $f$ ranging from 1 (that is, no dead zone) to 1000. In this case, the
maximum viscous timescale in the disc is initially at the outer edge
of the dead zone and is indicated by the three vertical dashed lines
in Figure 2 (note that the case of $f=1000$ is outside the scale of the plot). Although the disc expands to the point that the viscous timescale
at the outer edge is similar to $\tau$, the system cannot enter a
quasi-steady state until $\tau$ exceeds this {\it maximum} value. The
inclusion of a dead zone thus has the effect of merely delaying the
onset of the quasi-steady state. Depending on the chosen parameters, the delay might be long enough that the disc enters into a quasi-steady state on a timescale longer than the observationally constrained lifetime of discs. In this case, we would expect the disc to accrete throughout its life with a dimensionless accretion parameter below unity.



\subsection{External photo-evaporation and other processes}
\label{sec:other}

Several factors may however prevent or modify the attainment of
a viscous quasi-steady state. First of all, depending on the nature of
angular momentum transport, there may be no surface density profile
that allows the condition that $\dot M$ is independent of $r$, for example if there are regions where the angular momentum transport is completely inhibited.
In this case the system will undergo a series of accretion bursts due to gravo-thermal instabilities \citep{Armitage2001}; it would then spend the majority of its time in a state
of low accretion where $\eta \ll 1$ \citep{Jones2012}. Secondly,
disc evolution may be subject to other mass and angular momentum sinks.
Recent work \citep[e.g.,][]{Suzuki2009,BaiStone2013,Fromang2013,Simon2013} has suggested that disc secular evolution may
be strongly influenced by magnetically driven winds. The consequences
of such winds for disc secular evolution however depend on the mass loss
profile and, in particular, on how the ratio of the specific angular momentum
removed by the wind to the local Keplerian value (the so-called dimensionless lever arm of the wind) depends on radius.
Currently these  radial dependences are poorly constrained \citep{Armitage2013,Bai2016} and we do not
explore this scenario further here. On the other hand, disc thermal
photoevaporation is reasonably well understood, whether it derives from 
internal or external radiation sources \citep{Clarke2001,Owen2010,Johnstone1998,Adams2004,Clarke2007,Anderson2013,Facchini2016} and in this case 
angular momentum and mass are removed in a ratio that is given by the
local Keplerian specific angular momentum. \citet{Jones2012} demonstrated that
internal photoevaporation (which drives inside-out clearing) results in
$\eta < 1$ (i.e. \textit{inhibits} accretion). We 
now turn to investigate the effect of external photo-evaporation.

\subsubsection{External photo-evaporation}
\label{sec:external}

\begin{figure}
\includegraphics[width=1.05\columnwidth]{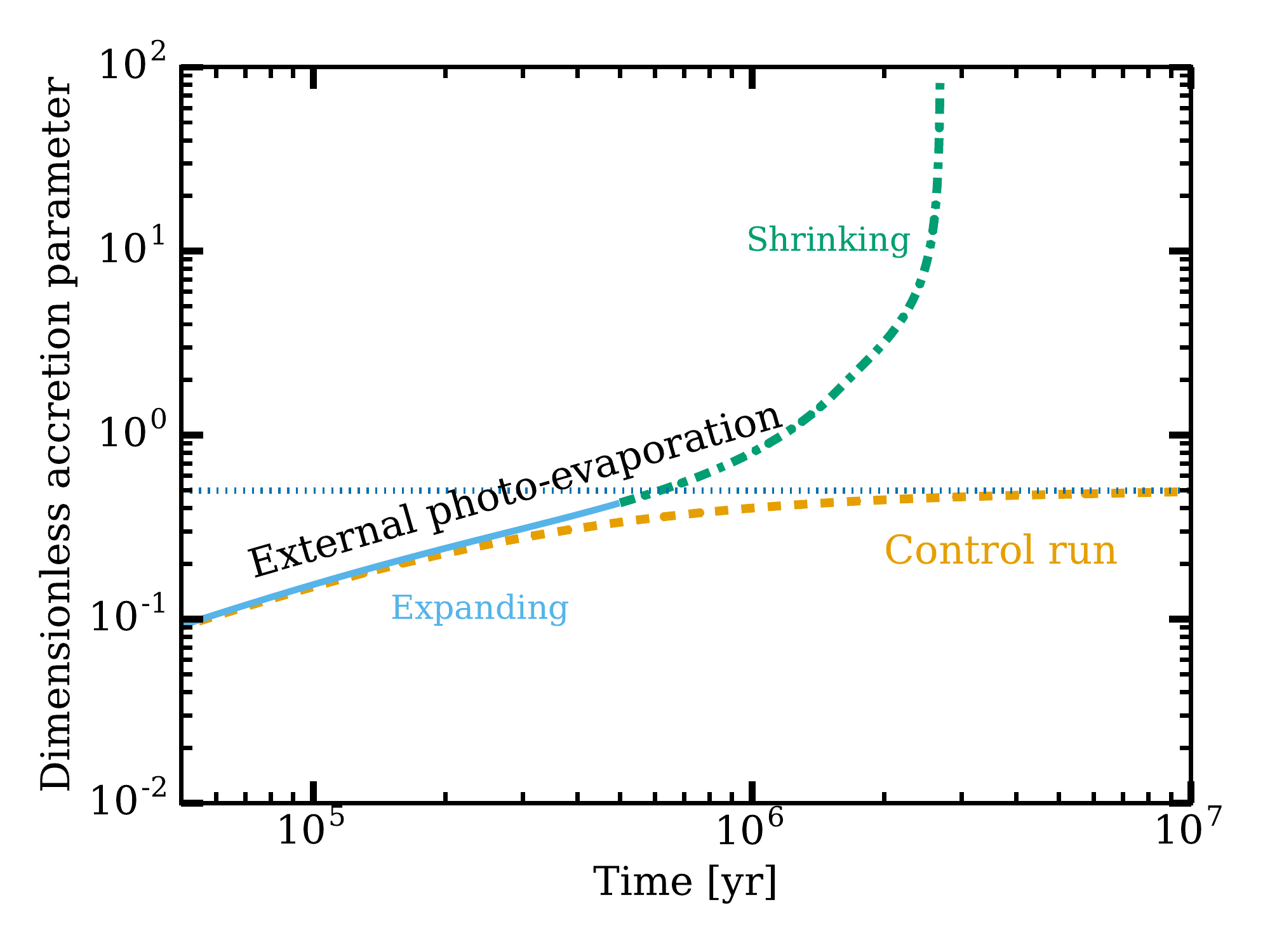}
\caption{Dimensionless accretion parameter of an externally photo-evaporating disc. The light blue solid line marks the expanding phase of the disc and the green dotted dashed one the shrinking one. Shown on the plot (orange dashed line) there is also the result from a control run where we have not included external photo-evaporation and the reference dimensionless accretion parameter of 0.5 (horizontal blue dotted line). Notice how only in the shrinking phase external photoevaporation starts to matter; during this phase the dimensionless accretion parameter becomes noticeably higher than unity. }
\label{fig:external}
\end{figure}

In Figure \ref{fig:external} we show the case of a disc subject to external
photoevaporation (light blue solid and green dotted dashed lines), compared with a control simulation without
photoevaporation (orange dashed line). The chosen parameters are such that the disc initially expands
viscously as normal (light blue part of the time evolution), but then reaches a radius where the
external mass loss rate is comparable with the accretion rate \citep{Clarke2007,Anderson2013}. At this
point, the disc is cleared from the outside in (green part of the time evolution); the mass of the disc quickly decreases as
the dominant mass reservoir in the disc (at large radii) is eroded. As the disc shrinks, the viscous time at the outer edge of the disc decreases because $t_\nu$ is an increasing function of radius (see Section \ref{sec:viscous}). This leads to a progressive increase in $\eta$ during the clearing phase. We expect other processes that truncate the disc, for example dynamical encounters, to produce a similar effect. In the case of a dynamical encounter, however, the disc will revert back to a dimensionless accretion parameter of unity on a viscous time-scale. Summarising, the evolution is distinct in two phases, one characterised by $\eta \sim 1$ where the effects of external photo-evaporation are almost negligible, and one characterised by $\eta \gg 1$ during the outside in clearing.

\begin{figure}
\includegraphics[width=\columnwidth]{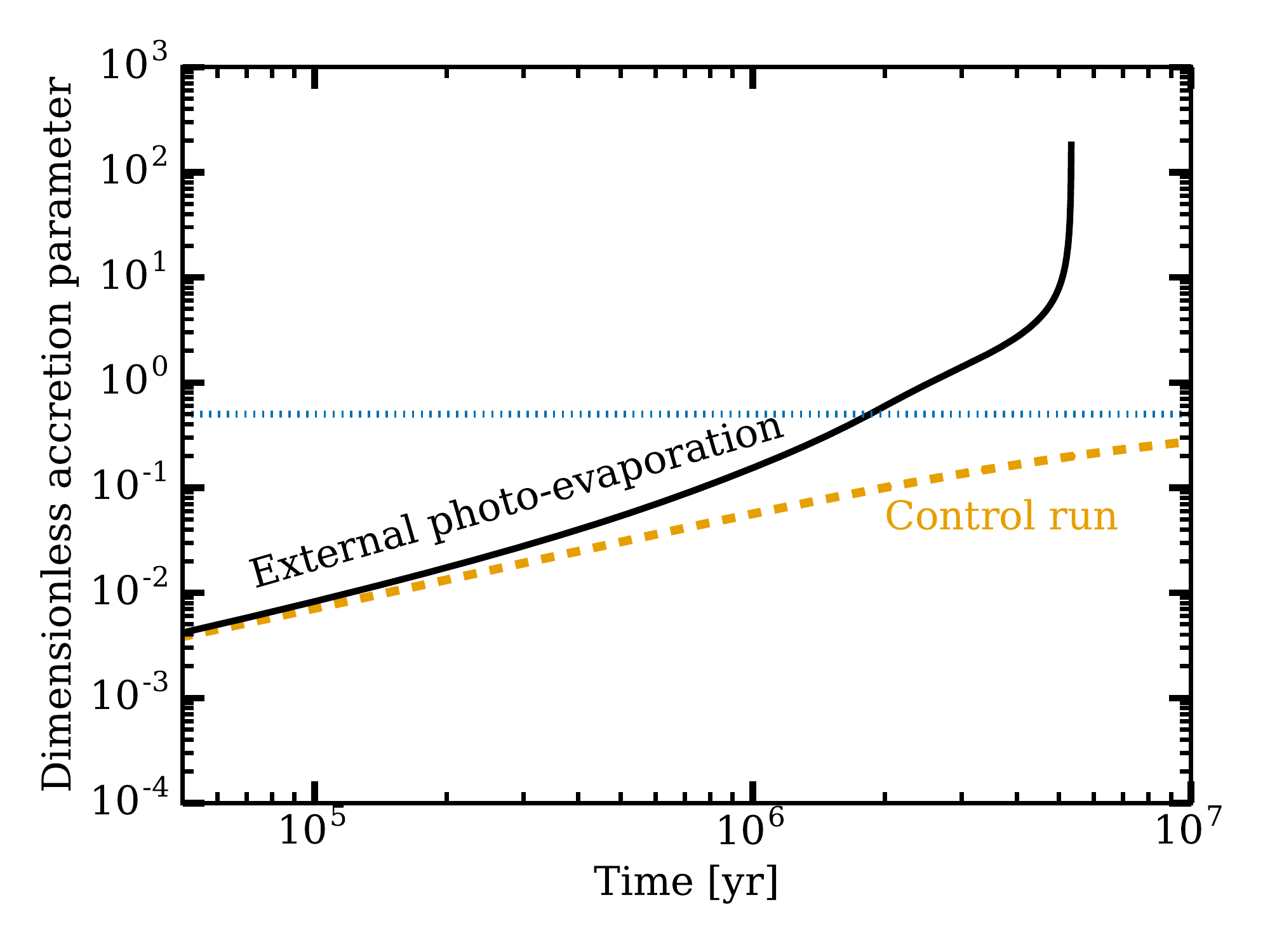}
\caption{Same as Figure \ref{fig:external} but for a disc that only experiences the shrinking phase. Solid black line: photo-evaporating disc. Orange dashed line: control run without photo-evaporation. Horizontal blue dotted line: the reference dimensionless accretion parameter of 0.5.}
\label{fig:externalbig}
\end{figure}

In Figure \ref{fig:externalbig}, the initial conditions involve instead a case where
external photoevaporation is important in controlling the disc radius
from the beginning and hence $\eta$ increases monotonically 
throughout the evolution. In this case the disc always accretes efficiently.

We thus conclude that external photo-evaporation enhances the disc dimensionless accretion parameter above unity during the outside in dispersal phase (i.e. at the end of the disc lifetime), but that it is only \textit{markedly} above unity during the very brief final clearing phases (see Figure \ref{fig:external}).

\subsection{Summary of theoretical expectations}

\begin{figure}
\includegraphics[width=\columnwidth]{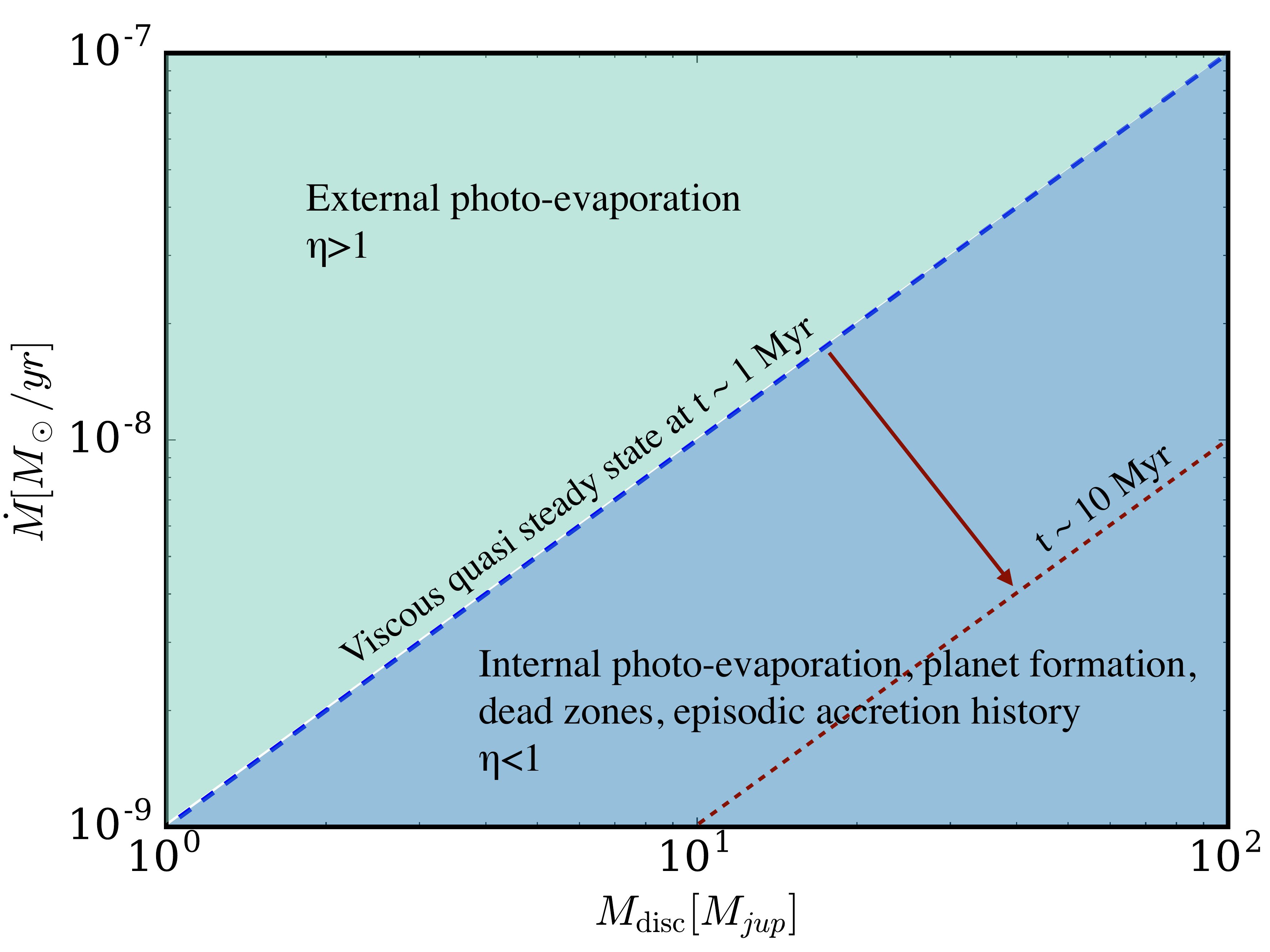}
\caption{Cartoon showing how the $M/\dot{M}$ ratio can inform us on the processes governing disc evolution. To draw the different regions we have assumed an age of the system; for reference we used 1 Myr, which is the typical age of observed proto-planetary discs, and indicated on the plot how the regions would move for an age of 10 Myr.}
\label{fig:schema}
\end{figure}

Combining the results of our numerical experiments with those
of \citet{Jones2012} we conclude that viscous evolution generally
results in $\eta \lesssim 1$. The case $\eta \sim 1$
corresponds to the establishment of  a viscous quasi-steady state; $\eta < 1$
can result from the system age being younger than the maximum
viscous timescale in the disc, from the fact that the
viscosity prescription does not admit a quasi-steady state
solution (episodic accretion\footnote{While episodic accretion produces high accretion efficiencies during the accretion burst phase, the disc spends most of its time accreting with low efficiencies \citep{Jones2012}.}) or else from internal photoevaporation. The {\it only}
scenario for which $\eta \gg 1$ is expected is that of
external photoevaporation during the final stages of disc dispersal\footnote{Dynamical encounters can have a similar effect as well as mentioned in Section \ref{sec:external}, but in practice photoevaporation becomes more important at substantially lower densities \citep{ScallyClarke2001}.}. We summarise these results in Figure \ref{fig:schema}.

\section{Comparison with observations}
\label{sec:observations}

In this section we gather existing data from the literature and analyse them in light of the theoretical considerations presented in the previous section. Note that, to compare to observations, we now plot accretion rate and disc mass, which are the quantities that the observations attempt to assess. Computing a dimensionless accretion parameter requires choosing an age of the system. In contrast to \citet{Jones2012}, in this work we do not use the individual ages of each star, as they are notoriously unreliable (e.g. \citealt{Soderblom2014}) but instead indicate in the plots the lines along which the dimensionless accretion parameter is unity for various assumptions about the stellar age. The typical ages of star forming regions containing proto-planetary discs are in the range 1-10 Myr \citep{Fedele2010}, although they might be underestimated by a factor of 2 according to recent estimates \citep{Bell2013}.

\subsection{The validity of the available tracers of the disc mass}

\begin{figure}
\includegraphics[width=\columnwidth]{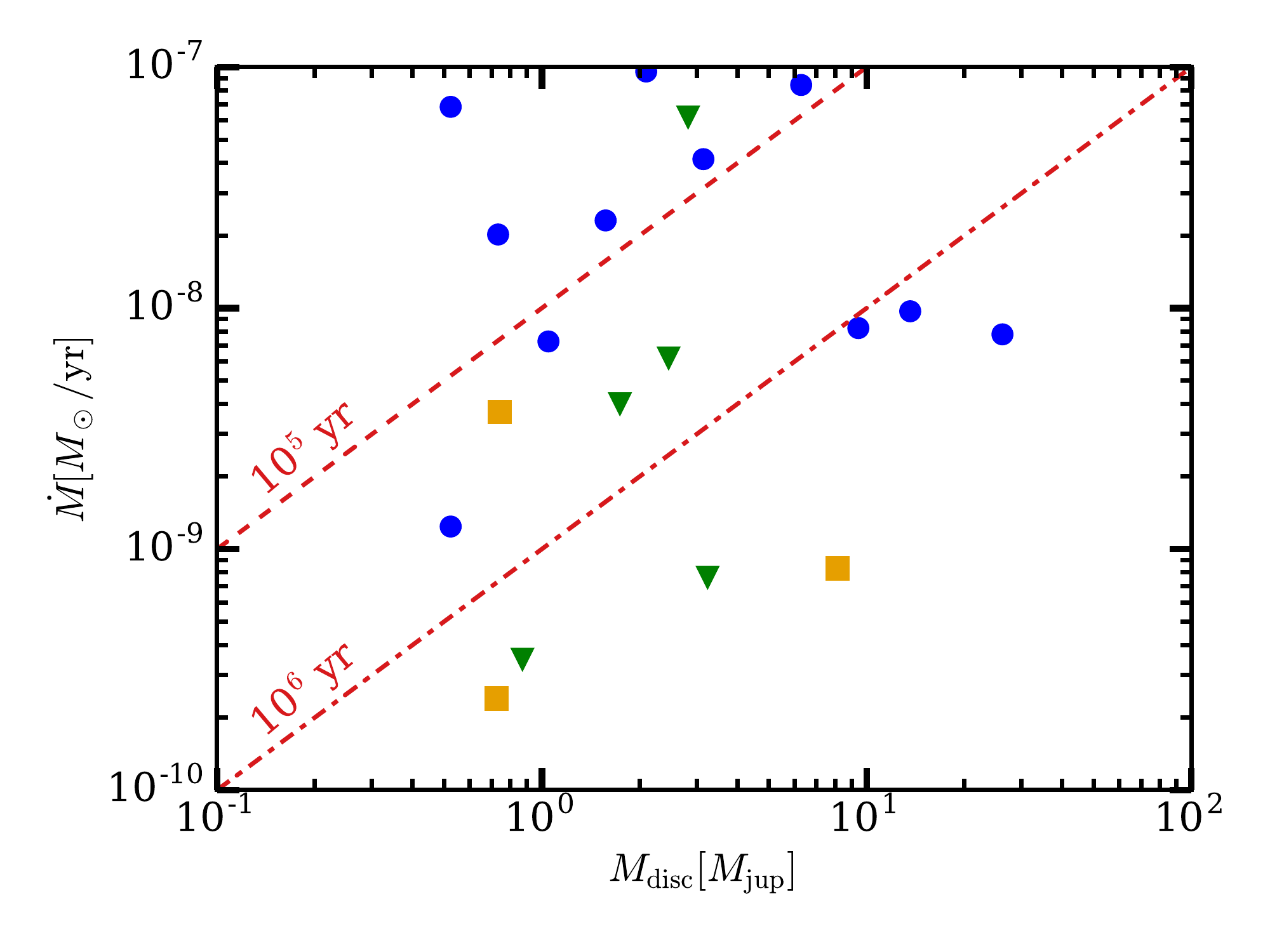}
\caption{Measurements of accretion rate and disc masses (deduced from the \textit{gas}) for all the sources where they are available. Blue points are for the sample of \citet{WilliamsBest2014}, green triangles for Lupus \citep{Ansdell2016}, orange squares for the objects with low accretion from the Lupus sample. The dashed and dotted dashed lines correspond to accretion efficiencies of unity assuming that the typical age is $10^5$ and $10^6$ years respectively.}
\label{fig:gas_masses}
\end{figure}

\begin{table*}
\caption{Values of the accretion rate collected from the literature for the sample of \citet{WilliamsBest2014}.}
\label{tab:willbest}
\begin{tabular}{cccc}
\toprule
Object &  Disc mass [$M_\odot$] &  $\dot{M} \ [M_\odot \mathrm{yr}^{-1}]$ & Reference\\
\midrule
AA Tau &           1.5e-03 &                               2.3e-08  &  \citet{Ingleby2013}  \\
CI Tau &           3.0e-03 &                               4.2e-08 & \citet{McClure2013} \\
CY Tau &           1.0e-03 &                               7.3e-09 & \citet{Hartmann98} \\
DL Tau &           5.0e-04 &                               6.8e-08 & \citet{White2001} \\
DO Tau &           2.0e-03 &                               9.6e-08 & \citet{Hartmann98}\\
Haro 6-13 &           6.0e-03 &                               8.4e-08 & \citet{Salyk2013}\\
IQ Tau &           7.0e-04 &                               2.0e-08 & \citet{Hartmann98} \\
TW Hydra &           5.0e-04 &                               1.2e-09 & \citet{Manara2014} \\
DM Tau &           9.0e-03 &                               8.3e-09 & \citet{Manara2014} \\
GG Tau &           1.3e-02 &                               9.7e-09 & \citet{Hartmann98}\\
IM Lup &           2.5e-02 &                               7.8e-09 & \citet{Alcala2016} \\
\bottomrule
\end{tabular}
\end{table*}

As mentioned in Section \ref{sec:intro}, observations do not directly yield the disc masses. 
By comparing the theoretical expectations with the observational data available from the surveys, we will examine how arguments developed in the previous section can be used to assess the reliability of the two main tracers of disc mass.

\subsubsection{Disc masses obtained via CO measurements}

Thanks to recent improvements in sensitivity to sub-mm emission lines, it is now possible to assess \textit{gas} masses through the observation of the rare, faint CO isotopologue C$^{18}$O. Thermo-chemical modelling \citep[e.g.,][]{Miotello2016} is necessary because the abundance relative to molecular hydrogen depends on photo-dissociation and freeze-out of the molecule. We collected data from the Lupus survey (\citealt{Ansdell2016} for the masses and \citet{Manara2016} for the accretion rates) and from the sample reported in \citet{WilliamsBest2014} (mostly consisting of sources from Taurus). For the latter sample, we surveyed the literature to find accretion rates (see Table \ref{tab:willbest}). We used the \citet{Siess2000} stellar evolutionary models in the interest of uniformity of calculation of accretion rates.

The data points show a low value of the $M_\mathrm{disc}/\dot{M}$ ratio (see Figure \ref{fig:gas_masses}). In particular, 12 out of the 19 data-points have a $M_\mathrm{disc}/\dot{M}$ ratio smaller than $10^6$ years, and 7 even smaller than $10^5$ years. For comparison, the age of Taurus is estimated to be 1-2 Myr \citep{Briceno2002,Luhman2003}. Note that in the plot we report only discs with C$^{18}$O detections. The sample is thus biased towards \textit{large} disc masses (i.e. bright discs), so that the ratio for the full sample could be even smaller. Moreover, if the recent revision of stellar ages by a factor of 2 \citep{Bell2013} is correct, the problem would be even more serious. 

Figure \ref{fig:gas_masses} demonstrates that the distribution
of points in the $M_{disc}, \dot M$ plane shows considerable
scatter and is incompatible with all objects having a single
dimensionless accretion parameter.
We have seen in Section
\ref{sec:expectations} that there are a variety of effects that can result in a
dimensionless accretion parameter of less than unity;  thus the most troubling
points are those with high accretion rates for which the
dimensionless accretion parameter is much larger than unity for any reasonable
stellar age estimate. In Section \ref{sec:expectations} we argued that a high
dimensionless accretion parameter is only to be expected in discs affected by external photo-evaporation. However, we find this explanation unlikely. Firstly, neither Taurus nor Lupus contains massive OB stars able to drive large photo-evaporative flows. Secondly, even if external photo-evaporation were eventually important for the evolution of these discs, as suggested by the recently revised photo-evaporation rates by \citet{Facchini2016}, the arguments presented in Section \ref{sec:external} show that a dimensionless accretion parameter significantly greater than unity can be sustained only at the end of the disc lifetime, during the outside in dispersal. We would then be in the uncomfortable situation in which we are seeing these discs close to the end of their lifetime. This is unlikely to hold true in different star forming regions.

The explanation that we favour is that gas masses based on CO isotopologues represent a substantial under-estimate of the true disc mass. Similar interpretations have been proposed in the astro-chemical community. For example, a comparison of masses derived from the CO and HD molecules in the nearby disc TW Hya \citep{Bergin2013} has led \citet{Favre2013} to conclude that CO is substantially less abundant (by a factor ranging $10^{-2}-0.3$) compared with what normally assumed. \citet{Kama2016} obtained a similar large under-abundance of both C and O in the same system by simultaneous fitting of emission lines of HD and different C and O bearing species. \citet{Miotello2017}, intepreting the relatively low CO based masses, suggested that carbon could be sequestered from CO to more complex molecules, or locked up in larger bodies. \citet{McClure2016}, again from HD measurements, found that DM Tau shows a carbon depletion up to a factor of 5 and GM Aur up to two orders of magnitude.

Here we demonstrate, for a sample much larger than that for which HD measurements are available, that available accretion rate data strongly disfavours the use of CO isotopologues data for the estimation of disc mass.

\subsubsection{Disc masses obtained from the dust}

We now  assess the reliability of disc masses based
on dust continuum measurements. This method has been widely used
since the seminal work of \citet{Beckwith1990} although it
is widely recognised that there are uncertainties associated with the unknown dust to gas ratio, the disc temperature, the grain opacity
and the effect of finite optical depth. The blue points
in Figure \ref{fig:onc} replot part of the data presented in \citet{Manara2016} for
the Lupus region\footnote{We concentrate here only on the most massive discs for comparison with the ONC.} using dust derived masses.
As noted by \citet{Manara2016},  there is a roughly linear
correlation between dust mass and accretion rate with a  relatively
modest scatter; this implies
a relatively narrow range of accretion efficiencies. Moreover the implied
median dimensionless accretion parameter is unity
for a plausible system age of  $\sim$ 2.3 Myr, compatible with the age of the region \citep{Alcala2014}.

It therefore appears that dust based mass estimates in Lupus\footnote{In the appendix we show that also for the sample of \citet{WilliamsBest2014}, used in the previous section, one recovers a dimensionless accretion parameter of roughly unity when using the masses derived from the dust, albeit with significantly more scattering than for Lupus sample.} are broadly
consistent with the theoretical expectations laid out in Section \ref{sec:expectations} and
\citet{Jones2012}. The relatively small scatter moreover
suggests that discs have evolved into a viscous quasi-steady state. Note that this consistency is achieved assuming a
canonical dust to gas ratio of 1:100, suggesting that the true value does
not deviate from this by a large factor more than a factor $\sim$ 2.

\subsection{Evidence of external photo-evaporation in Orion}

\begin{figure}
\includegraphics[width=\columnwidth]{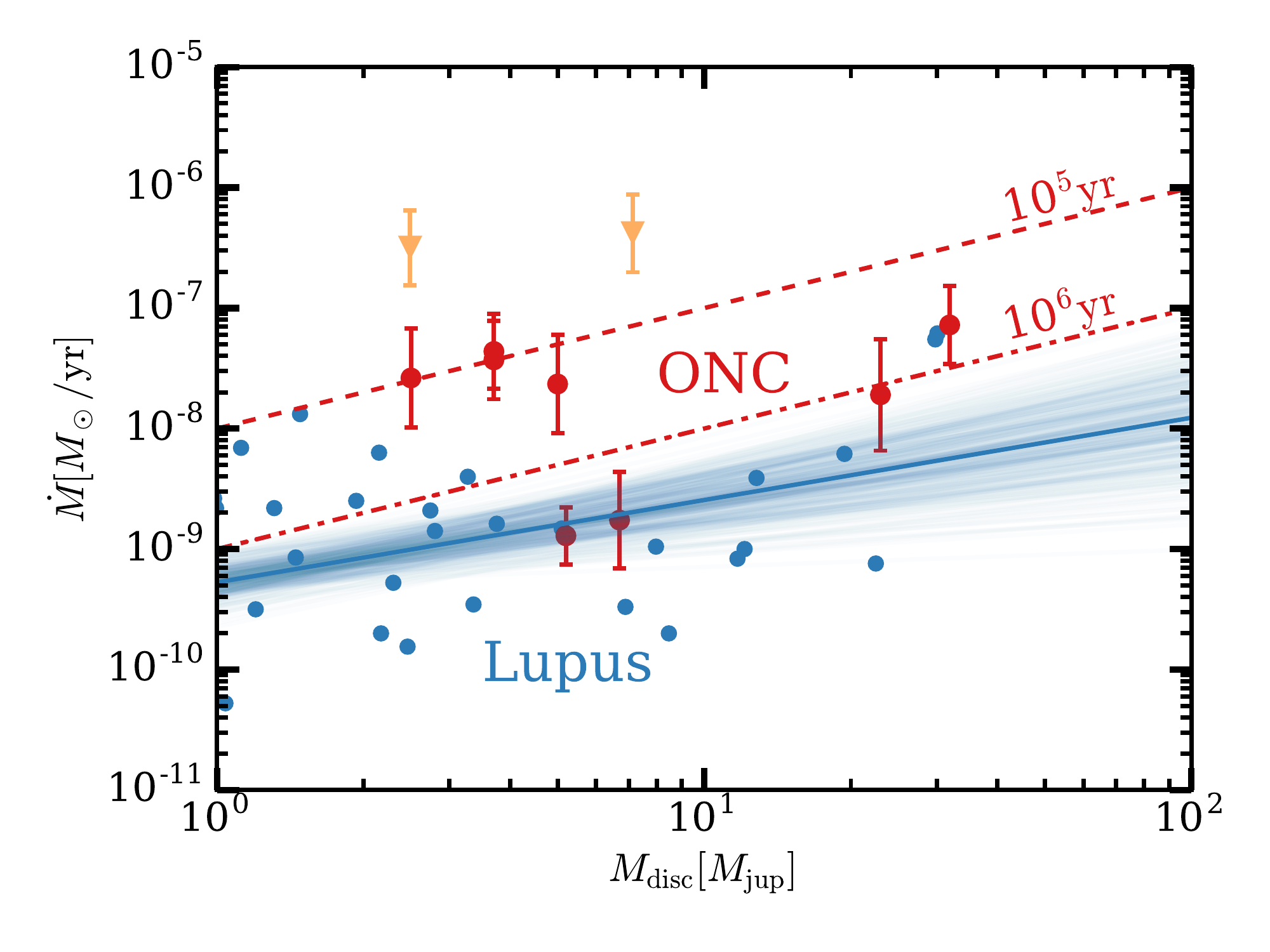}
\caption{Measurements of accretion rate and disc masses (deduced from the \textit{dust}) for objects in the Lupus cloud as blue points \citep{Manara2016}. The blue solid line shows the best fit to the data, and in light blue we plot a subsample of the results of some Markov Monte-Carlo chains to show the uncertainty in the correlation. The objects in the ONC are shown as red points. The error bars take into account the error on the mass accretion rate propagated from the photometric error. The orange triangles denote objects that show some nebulosity in the HST images, for which the measurements of the accretion rate are more uncertain. Despite the relatively small size of the sample, it can be seen how the $M/\dot{M}$ ratio is smaller than the age of the system ($\sim$ 1 Myr, shown by the dotted dashed red line), which we interpret as due to the effects of external photo-evaporation.}
\label{fig:onc}
\end{figure}

After having shown in 3.1.2 that dimensionless accretion parameter
data in Lupus (using  {\it dust} based mass estimates)
 is compatible with the generic expectations of evolutionary models, we need to further investigate whether
dust based masses are sufficiently {\it accurate} for the
dimensionless accretion parameter to be useful in diagnosing
possible differences between disc populations. With this
in mind we make a comparative study of
dimensionless accretion parameter data in a low mass star forming
region (Lupus) with that obtained in the Orion Nebula Cluster.
We have abundant observational evidence for the importance
of external photoevaporation in the ONC \citep[e.g.,][]{1987ApJ...321..516C,Johnstone1998}, notably
in the form of the so-called proplyds \citep{Odell93,2005A&A...441..195V}: discs associated with an offset ionization front. They are commonly interpreted as discs losing mass through a wind driven by the FUV photons of $\theta_1$C, the most massive star in the ONC; the ionization front traces the location where the wind becomes optically thin to EUV photons. Given the
modelling in Section 2, we would thus {\it expect} accretion
efficiencies to be significantly higher in the ONC since outside in-clearance reduces disc mass more than accretion rate. If this
turns out to be the case, the dimensionless accretion parameter data
can then be used as a diagnostic of the importance of
external photoevaporation. This would enable the diagnostic
to be used in future in analysing
clusters where the proplyd population
cannot be well characterised, either because of distance or because external photoevaporation drives a significative mass loss even if does not produce prominent proplyds.

We have collected dust masses in the ONC from the recent ALMA measurements of \citet{Mann2014} and \citet{Eisner2016}. We then cross-correlated the catalogue with the measurements of accretion rate from HST H$\alpha$ photometry presented in \citet{Manara2012}. This resulted in a sample of 11 objects with both $\dot{M}$ and disk mass measurements. Due to observational limitations, the sample is obviously not complete and is biased towards large disc masses and accretion rates, given that these sources are easier to detect. Figure \ref{fig:onc} shows as red points the position of these targets on the $\dot{M}-M_\mathrm{dust}$ plane, where $M_\mathrm{dust}$ is the disc mass calculated assuming a gas-to-dust ratio of 100. Despite the relatively small size of the sample, it can be seen how the $M/\dot{M}$ ratio is smaller than the age of the region ($\sim$ 2.2 Myr, \citealt{Reggiani2011}) in 8 out of 11 objects. To compare with Lupus, as mentioned in the previous section we have plotted with the blue line the best fit to the data in Lupus \citep{Manara2016} and in light blue a subsample of the results of some Markov Monte-Carlo chains to show the uncertainty in the correlation. It is clear how the distribution of efficiencies in the ONC is higher than in Lupus. 

Some doubt could be cast on measurements of $\dot{M}$ derived solely from H$\alpha$ in proplyds, as contamination from unresolved ionization fronts of proplyds or from the strong background is possible and could lead to an overestimate of $\dot{M}$. We inspected visually the HST images for these objects and we did not find evidence for resolved ionization fronts, but for two objects we notice a small increase of the nebulosity of the background around them. We plot them as orange triangles and consider them as more uncertain; we note however that the other datapoints (red points) are still above the relation of Lupus (where instead the datapoints are scattered on both sides of the best fit, see the Lupus datapoints shown in blue) even if we were to remove them. As a further check, we have computed the median accretion timescale $M/\dot{M}$ of the blue points which is $\sim 3\times 10^5 \ \mathrm{yr}$, a factor of $\sim$7 less than the age of the region.

We conclude that the inner region of the ONC is different from Lupus \citep{Ansdell2016,Manara2016}. In light of the arguments presented in Section \ref{sec:external}, this shows that the existing data is already of sufficient quality to recognise the fingerprint of external photo-evaporation in these discs. Indeed all of our sources (except one) lie inside a projected distance of 0.3 pc from $\theta^1$
Ori C, which is where external photo-evaporation is expected to be important \citep{StorzerHollenbach1999}. These discs are therefore currently being dispersed outside in, or, in other words, at the end of their life-time (see also discussion in \citealt{Gorti2016}).

\section{Conclusions}
\label{sec:conclusions}

In this paper we have shown how the availability of mass accretion rate and disc mass estimates for large samples of young stars provides an opportunity to both assess the reliability of these estimates
and also to use them to explore systematic effects in different
star forming regions. 

We have introduced the concept of dimensionless accretion parameter ($\eta$, see Eq. \ref{eq:eta})  as a diagnostic for distinguishing between various theoretical scenarios. Extending the work of \citet{Jones2012}, we have demonstrated the mapping between the value of $\eta$ and a broad range of evolutionary scenarios (see Figure \ref{fig:schema} for a schematic depiction). In particular we have confirmed that the dimensionless accretion parameter is expected to be around unity when disc evolution is driven by viscosity and the disc can reach a quasi steady-state. We have shown how this is a general result which does \textit{not} depend on idealised assumptions about the viscosity. While a variety of processes can \textit{lower} the dimensionless accretion parameter below unity, we have shown that external photo-evaporation is the only one capable of \textit{increasing} it and that moreover a significant raising
of dimensionless accretion parameter is only expected in advanced stages of
outside-in clearing of discs.

The generic expectations from various evolutionary scenarios can also be used to assess the validity of the existing diagnostics of the disc mass. When considering disc masses derived from C$^{18}$O in the weakly irradiated environments of Lupus and Taurus, we have shown how at face value the only way to interpret the current data is if the evolution of those discs would be dominated by external photo-evaporation. Because of the lack of massive stars in those regions, we reject this interpretation and propose that the current models used for interpreting CO data lead to substantial underestimates of the true disc gas mass, in line with similar findings based on detailed chemical modelling of small numbers of sources. Dust based gas estimates however result in accretion efficiencies
of around unity as expected, suggesting that the dust can be employed as a reasonable measure of the disc mass, with a value of the dust to gas ratio close (within a factor of $\sim$ 2) to the canonical ISM value of 100.

To show that the existing data (using dust based disc mass estimates) is already accurate enough to find differences between regions through dimensionless accretion parameter analysis, we have also perfomed this analysis in the ONC using existing observational data. These discs are subject to an intense ultraviolet field and are thus expected to drive considerable photo-evaporative flows which, as shown in Section \ref{sec:expectations}, should result in $\eta$ values considerably greater than unity. The available data suggests that this is indeed the case (Figure \ref{fig:onc}). The sample is however currently small and biased towards objects with large disc masses and
accretion rates. Further studies are required to substantiate the effect in larger samples. 

We also highlight that, if confirmed, it implies that dimensionless accretion parameter data can be used as a new diagnostic of the importance of environmental disc evaporation in star forming regions.

\section*{Acknowledgements}
We thanks an anonymous referee for comments that improved the clarity of the paper. This work has been supported by the DISCSIM project, grant
agreement 341137 funded by the European Research Council under
ERC-2013-ADG. CFM gratefully acknowledges an ESA Research Fellowship. We thank Phil Armitage, Jacob Simon, Tilman Birnstiel, Mario Flock, Leonardo Testi and Antonella Natta for inspiring discussions and Richard Booth and Mihkel Kama for providing comments on a draft version of this paper.




\bibliographystyle{mnras}
\bibliography{viscosity} 


\appendix
\section{Taurus dust masses}

\begin{figure}
\includegraphics[width=\columnwidth]{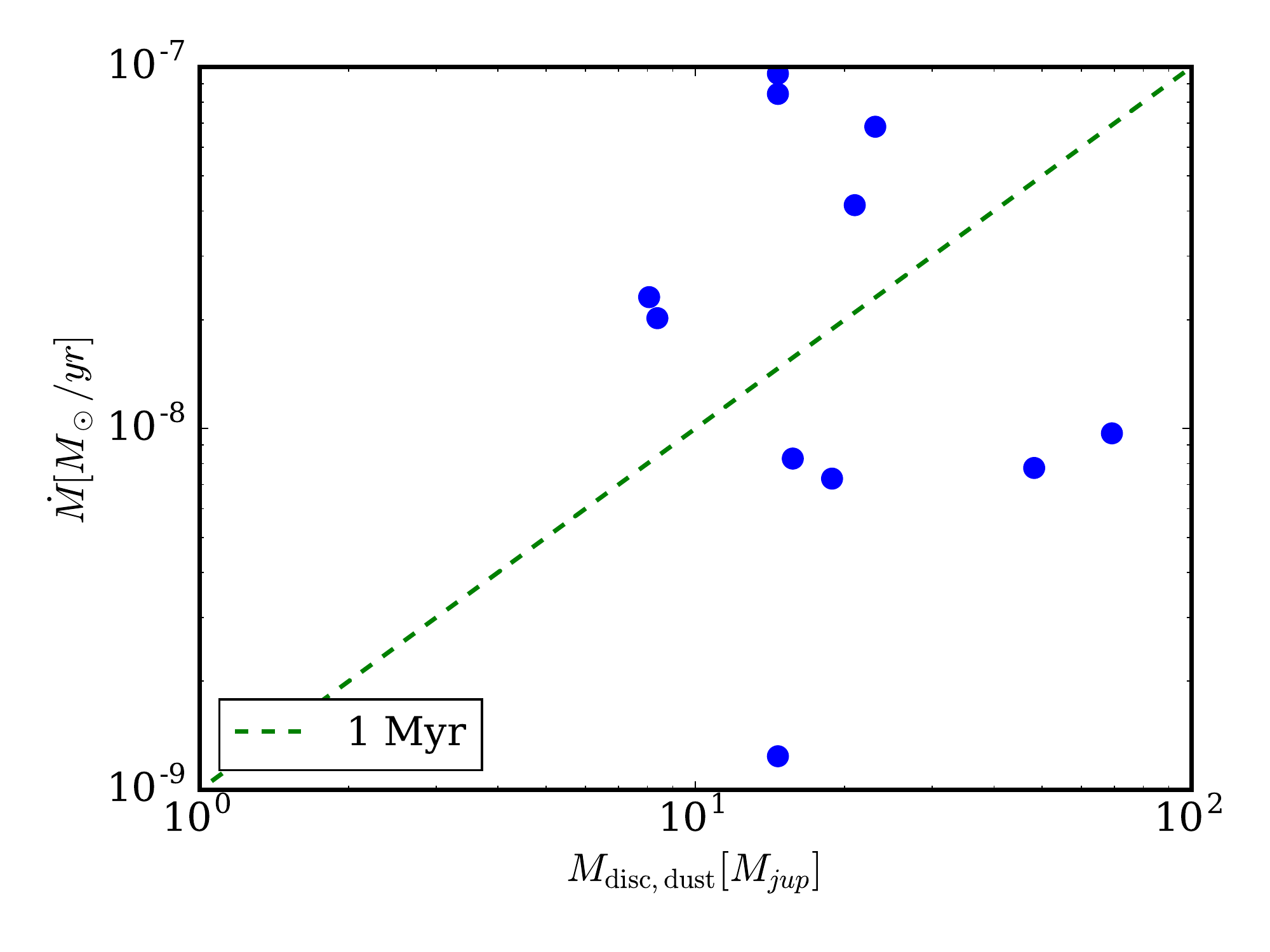}
\caption{Measurements of disc masses (derived from the \textit{dust}) and accretion rate for the sample of \citet{WilliamsBest2014}. The values are evenly spread around the 1 Myr line.}
\label{fig:taurus_dust}
\end{figure}

\citet{Manara2016} have shown in Lupus that the dimensionless accretion parameter is close to unity using the disc masses derived from the dust and using a standard dust to gas ratio of 100. Here we show that the same holds also for the sample of \citet{WilliamsBest2014}. We plot the results in Figure \ref{fig:taurus_dust}. The ratio of dust derived disc mass to accretion
rate shows considerable scatter but is not systematically offset from the
system age ($\sim$ a Myr). This indicates a dimensionless accretion parameter which is close to unity as found in Lupus. Note that the Lupus data have significant advantages compared to this sample: a higher number statistics, higher spectral resolution, better flux calibration and a homogeneous data analysis. This accounts for the much larger scatter in the Taurus
data.

\bsp	
\label{lastpage}
\end{document}